\documentclass{article}
\usepackage{spconf}
\usepackage{graphicx}
\usepackage{amsmath,amssymb,latexsym,amsthm,mathrsfs,bbm}
\usepackage{cite,algorithm,algorithmic,url}
\usepackage{multirow}
\usepackage{booktabs}
\usepackage{array}
\usepackage{threeparttable}
\usepackage{comment}
\newcolumntype{C}[1]{@{}>{\centering\arraybackslash}p{#1}@{}}

\title{APPLADE: Adjustable plug-and-play audio declipper \\ combining DNN with sparse optimization}

\name{Tomoro Tanaka$^\dag$, Kohei Yatabe$^\dag$, Masahiro Yasuda$^\ddag$, Yasuhiro Oikawa$^\dag$}
\address{$^\dag$ Department of Intermedia Art and Science, Waseda University, Tokyo, Japan\\
$^\ddag$ NTT Corporation, Tokyo, Japan\vspace{-10pt}}

\begin{document}
\ninept
\maketitle
\begin{abstract}
In this paper, we propose an audio declipping method that takes advantages of both sparse optimization and deep learning.
Since sparsity-based audio declipping methods have been developed upon constrained optimization, they are adjustable and well-studied in theory.
However, they always uniformly promote sparsity and ignore the individual properties of a signal.
Deep neural network (DNN)--based methods can learn the properties of target signals and use them for audio declipping.
Still, they cannot perform well if the training data have mismatches and/or constraints in the time domain are not imposed.
In the proposed method, we use a DNN in an optimization algorithm.
It is inspired by an idea called plug-and-play (PnP) and enables us to promote sparsity based on the learned information of data, considering constraints in the time domain.
Our experiments confirmed that the proposed method is stable and robust to mismatches between training and test data.%
\vspace{-4pt}
\end{abstract}

\vspace{-2pt}
\section{Introduction}
\label{sec:intro}
\vspace{-2pt}

\begin{figure}[t]
	\includegraphics[width=0.99\columnwidth]{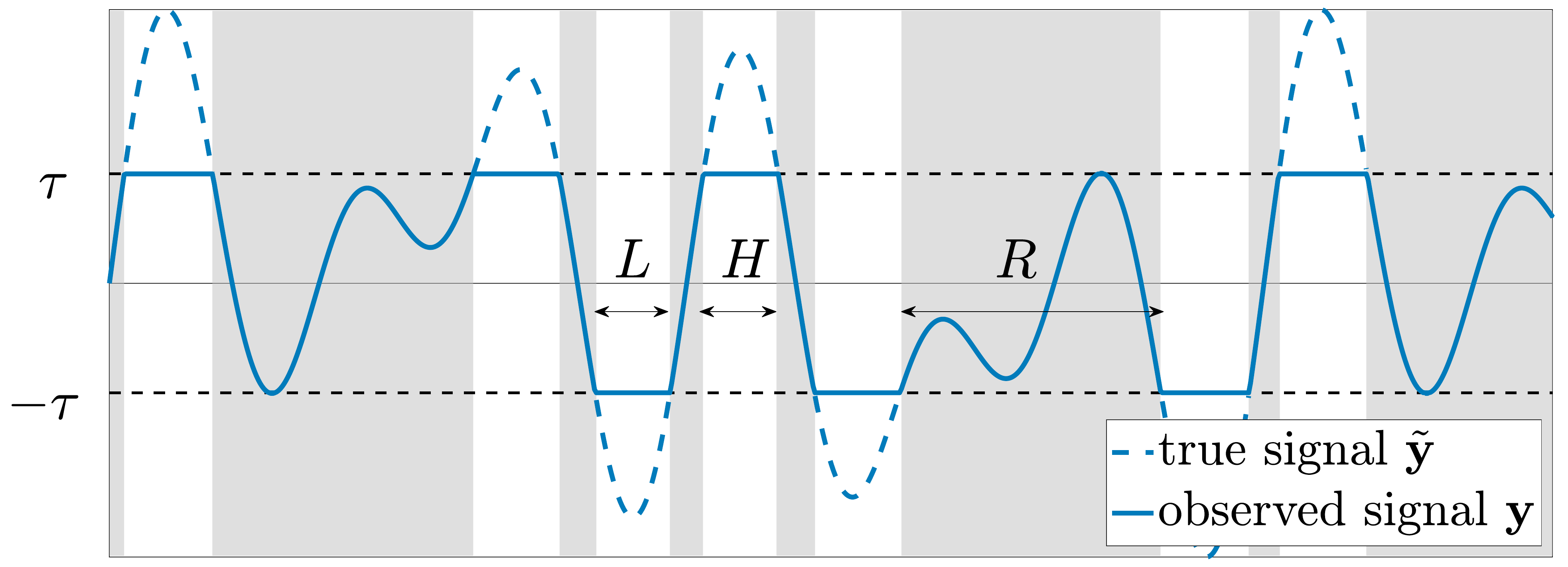}
	\vspace{-0.3cm}
	\caption{The true signal $\tilde{\bf y}$ and an observed clipped signal $\bf y$ with the clipping threshold $\tau$. Gray-filled areas indicate that clipping has no effect. $R$ is the set of the non-clipped indices, and $H$ and $L$ are the sets of the clipped indices above and below $\pm\tau$, respectively.}
	\label{fig:clip}
\end{figure}

The aim of audio declipping is to restore clipped signals.
Clipping is a common nonlinear distortion that occurs due to restriction of recording system and/or audio processing.
As illustrated in Fig.~\ref{fig:clip}, the samples of the true signal exceeding a dynamic range $[-\tau, \tau]$ with $\tau>0$ are truncated (so-called \textit{hard clipping}). 
Since clipping degrades audio quality \cite{q2,q1} and has bad effects on subsequent processing \cite{r2,r1}, audio declipping is desired.

Audio declipping has been realized by various methods \cite{IHT,SS,ASPADE,PWL1,multi,summary,DNN3,DF,TTF}.
Among them, sparsity-based methods \cite{IHT,SS,ASPADE,PWL1,multi,summary} have been studied for a long time.
They introduce an optimization problem consisting of a data-fidelity term and a sparsity-promoting term.
The data-fidelity term makes solutions feasible, and \textit{clipping consistency} \cite{summary} was considered in recent studies \cite{ASPADE,PWL1,multi}.
The sparsity-promoting term makes solutions sparse in the time-frequency (T-F) domain.
This is motivated by the fact that clipping produces extra harmonic components.
Sparsity-based methods are well studied in theory and have some adjustable parameters for variation in the properties of data.

Deep neural network (DNN)--based audio declipping methods have also been studied lately \cite{DNN3,DF,TTF}.
A DNN can learn the properties of data and perform audio declipping based on them.
In \cite{DF}, deep filtering was utilized, and complex-valued filters were estimated by a DNN composed of the bidirectional long short-term memory layers \cite{blstm}. 
In \cite{TTF}, the waveforms or magnitude spectrograms of non-clipped signals were directly estimated by a DNN which has the U-Net architecture \cite{unet}.
However, these methods cannot perform well when the properties of input data (such as scale and clipping threshold $\tau$) are greatly different from those of the training data.
Furthermore, the clipping consistency has not been explicitly considered during processing.

In this paper, we propose an audio declipping method that combines a DNN with sparse optimization and named it APPLADE (\textbf{a}djustable \textbf{p}lug-and-\textbf{pl}ay \textbf{a}udio \textbf{de}clipper). 
We use a DNN in an optimization algorithm to promote sparsity based on the learned properties of data, considering the clipping consistency.
Since the proposed method inherits the adjustability of the sparsity-based methods, it can handle a wide variety of data whose properties are greatly different from the training data.
The DNN and its usage are designed not to bother stability of an optimization algorithm and not to increase execution time so much.
Our experiments confirmed that the proposed method is stable and robust to differences between training and test data.

\vspace{-2pt}
\section{Preliminaries}
\label{sec:prelim} 
\vspace{-2pt}

In this paper, we consider hard clipping, in which signals' samples that exceed a certain dynamic range $[-\tau, \tau]$ are truncated.
Let $\tilde{\bf y}\in\mathbb{R}^T$ be the true signal, and ${\bf y}\in\mathbb{R}^T$ an observed clipped signal, 
\vspace{-2pt}
\begin{equation}
y[t]=\left\{
\begin{array}{cl}
    {\tau} &(\tilde y[t]\geq{\tau})\\
    \tilde y[t] &(-{\tau}<\tilde y[t]<{\tau}),\\
    {-\tau} &(\tilde y[t]\leq{-\tau})
\end{array} \right.
\label{eq:clip}
\vspace{-2pt}
\end{equation}
where $T$ is the length of the signal.
Our aim is to estimate the true signal $\tilde{\bf y}$ only from the clipped signal ${\bf y}$.
The situation is briefly shown in Fig.~\ref{fig:clip}.
As illustrated, their indices can be divided into three disjoint sets, $H=\{t\in[1,T]\;|\;{y[t]}\geq\tau\}$, $R=\{t\in[1,T]\;|\;|{y[t]}|<\tau\}$, and $L=\{t\in[1,T]\;|\;{y[t]}\leq-\tau\}$.

\vspace{-3pt}
\subsection{Sparsity-based audio declipping methods}
\label{ssec:sparse}
\vspace{-1.5pt}

Sparse optimization has been applied to audio declipping \cite{IHT,SS,ASPADE,PWL1,multi,summary}.
It is realized by solving the following optimization problem: 
\vspace{-2pt}
\begin{equation}
    \text{Find}\;\;{\bf{x}^\star}\;{\in}\;\arg\min_\mathbf{x}\:
\mathcal{S}({\mathcal G}{\bf x})\;\;\text{subject to}\;\;{\bf x}\in\Gamma,
\label{eq:optimization}
\vspace{-2pt}
\end{equation}
where the set of feasible signals $\Gamma$ is defined as
\vspace{-2pt}
\begin{equation}
{\Gamma}=\left\{{\bf{x}}\in\mathbb{R}^T\;\left|\;
\begin{array}{ll}
    {x[t]}\geq\tau & (t\;{\in}\;H) \\
    {x[t]}={y[t]} & (t\;{\in}\;R) \\
    {x[t]}\leq-\tau & (t\;{\in}\;L) \\
\end{array}
\right\}\right.,
\vspace{-2pt}
\end{equation}
$\mathcal{S}$ is a sparsity-promoting function (including $\ell_1$-norm, weighted $\ell_1$-norm \cite{PWL1}, $\ell_0$-norm \cite{IHT,ASPADE}), $\mathcal{G}$ is the discrete Gabor transform (DGT),
\vspace{-2pt}
\begin{equation}
    ({\mathcal{G}}{\bf x})[m,n] = \sum^{T-1}_{t = 0}x[t + an]\,g[t]\,e^{-2{\pi}{\mathrm{i}}mt/M},
\label{eq:DGT}
\vspace{-2pt}
\end{equation}
${\bf g}\in{\mathbb R}^T$ is a window, $\mathrm{i}$ is the imaginary unit, $a$ denotes the time shifting step, $n\in[0,N-1]$ and $m\in[0,M-1]$ are the time and frequency indices, respectively, satisfying $aN = T$, and the indices are to be understood modulo $T$.
From Eq.~\eqref{eq:clip}, $x^\star[t]$ should be exactly same as $y[t]$ for all $t \in R$.
Moreover, it must be greater than $\tau$ for all $t \in H$ and smaller than $-\tau$ for all $t \in L$.
That is, every solution must coincide with $\bf{y}$ after applying the same clipping.
This is called clipping consistency \cite{summary}, and the constraint in Eq.~\eqref{eq:optimization} forces ${\bf x}^\star$ to satisfy this consistency.
The sparsity-promoting term $\mathcal{S}({\mathcal{G}{\bf x}})$ makes a solution sparse in the T-F domain.
Since clipping produces extra harmonic components, promoting sparsity in the T-F domain will remove them and make a solution closer to non-clipped signals.

\vspace{-5pt}
\subsection{DNN-based methods}
\label{ssec:DNN}
\vspace{-1pt}

DNN-based audio declipping methods have begun to be studied lately \cite{DNN3,DF,TTF}.
A DNN approximates a mapping that estimates the true signal only from the features of a clipped signal.
Let $\mathcal{F}^{\text{TFD}}$ and $\mathcal{F}^{\text{TD}}$ be mappings in the time domain and T-F domain, respectively.
In \cite{DNN3}, the target mapping was $\mathcal{F}^{\text{TFD}}(\text{mel}(\mathcal{G}{\bf y})) = \text{mel}(\mathcal{G}\tilde{\bf y})$, where $\text{mel}(\cdot)$ extracts the mel-frequency cepstrum coefficients features.
In \cite{DF}, a DNN was trained to estimate the complex-valued deep filters.
In \cite{TTF}, both mapping $\mathcal{F}^{\text{TFD}}(\text{log}(|\mathcal{G}{\bf y}|^2)) = \text{log}(|\mathcal{G}\tilde{\bf y}|^2)$ and $\mathcal{F}^{\text{TD}}({\bf y}) = \tilde{\bf y}$ were approximated, where $|\cdot|$ is the element-wise absolute value.
DNN-based methods can restore non-clipped signals based on the learned properties of training data.
%However, it can be difficult for these methods to perform well when the features of test data, such as amplitude or/and $\theta_c$, are greatly different from those of training data.
%Moreover, the clipping consistency is considered during neither training nor testing.
%These disadvantages can result in failure of restoring desired non-clipped signals.

%\subsection{Plug-and-Play}
%\label{ssec:PnP}

%Plug-and-Play (PnP) is one of the methods to utilize a DNN in an iterative optimization algorithm to solve optimization problems, such as Eq.~\eqref{eq:optimization}.
%PnP has been studied mainly in the field of image processing \cite{PnP1,PnP2,PnP3,PnP4,PnP5,PnP6}.
%A DNN is inserted into an iterative algorithm and is expected to act as a proximity operator \cite{proximal,playwdual,monotone}.
%This idea stems from the fact that the proximity operator of a function $f$ can be viewed as a Gaussian denoising operator based on the prior distribution $\text{exp}(-f(\cdot))$ \cite{PnP4,DeGLI}.
%PnP methods have been studied because of its competitive effectiveness thanks to a lot of advantages from the both frameworks.

\vspace{-2pt}
\section{Proposed method: APPLADE}
\label{sec:prop}
\vspace{-2pt}

In this paper, we propose an audio declipping method named APPLADE.
It is inspired by the plug-and-play (PnP) method \cite{PnP1,PnP2,PnP3,PnP4,PnP5,PnP6}, which utilizes a DNN in an optimization algorithm.

\vspace{-5pt}
\subsection{Alternating Direction Method of Multiplier (ADMM)}
\vspace{-0.5pt}

In this paper, we use ADMM \cite{proximal} to solve Eq.~\eqref{eq:optimization}.
By applying ADMM to Eq.~\eqref{eq:optimization}, we obtain the following iterative procedure:
\vspace{-2pt}
\begin{align}
    {\bf x}^{[k+1]} &= \Pi_\Gamma(\mathcal{G}^*({\bf v}^{[k]} - {\bf u}^{[k]})),\label{eq:1}\\
    {\bf v}^{[k+1]} &= \text{prox}_{(1/\rho){\mathcal {S}}}(\mathcal{G}{\bf x}^{[k+1]} + {\bf u}^{[k]}),\label{eq:2}\\
    {\bf u}^{[k+1]} &= {\bf u}^{[k]} + \mathcal{G}{\bf x}^{[k+1]} - {\bf v}^{[k+1]},\label{eq:3}
    \vspace{-2pt}
\end{align}
where $\rho\in\mathbb{R}_{+}$, ${\bf v}\in\mathbb{C}^{M{\times}N}$ and ${\bf u}\in\mathbb{C}^{M{\times}N}$ are the auxiliary variables, $\mathcal{G}^*$ is the adjoint of $\mathcal{G}$ assuming the operator is a tight Parseval frame \cite{ASPADE}, $\Pi_\Gamma$ is the projection operator onto $\Gamma$ \cite{PWL1,summary},
\vspace{-2pt}
\begin{equation}
    (\Pi_{\Gamma}({\bf {x}}))[t]=
    \left\{\begin{array}{cl}
    \text{max}(x[t],\tau)& (t\;{\in}\;H)\\
    y[t] & (t\;{\in}\;R)\\
    \text{min}(x[t],-\tau) & (t\;{\in}\;L)
\end{array} \right.,
\label{eq:pj}
\vspace{-2pt}
\end{equation}
%Eq.~\eqref{eq:1} is easily solved by utilizing the proximity operator, %\cite{proximal,playwdual,monotone}, defined as
%\begin{equation}
	%\text{prox}_{{\mu}f}({\bf x}) = %\text{arg}\;\underset{{\bf %y}}{\text{min}}\Bigl[\,f({\bf %y})+\frac{1}{2\mu}\bigl\|{\bf x}-{\bf %y}\bigr\|_2^2\,\Bigr],
	%\label{eq:prox}
%\end{equation}
%for a function $\mu f$ and $\mu\in\mathbb{R}_+$. 
%\begin{equation}
    %{\bf x}^{[k+1]} = \Pi_\Gamma(\mathcal{G}^*({\bf v}^{[k]} - {\bf u}^{[k]})),
    %\label{eq:re1}
%\end{equation}
%where $\mathcal{G}^*$ is the adjoint of $\mathcal{G}$ assuming the operator is a tight Parseval frame \cite{ASPADE}, and $\Pi_\Gamma$ is the projection operator onto $\Gamma$ \cite{PWL1,summary},
%\begin{equation}
    %(\Pi_{\Gamma}({\bf {x}}))[l]=
   % \left\{\begin{array}{cl}
    %\text{max}({\bf x}[l],{\theta}_c)& (l\;{\in}\;H)\\
    %{\bf x}[l] & (l\;{\in}\;R)\\
    %\text{min}({\bf x}[l],-{\theta}_c) & (l\;{\in}\;L)
%\end{array} \right..
%\label{eq:pj}
%\end{equation}
%Likewise, Eq.~\eqref{eq:2} is rewritten as follows:
%\begin{equation}
    %{\bf v}^{[k+1]} = \text{prox}_{(\lambda/\rho){\mathcal {S}}}(\mathcal{G}{\bf x}^{[k+1]} + {\bf u}^{[k]}),
    %\label{eq:re2}
%\end{equation}
and $\text{prox}_{(1/\rho)\mathcal{S}}$
is the proximity operator \cite{proximal,playwdual,monotone}.
For example, the proximity operator of $\ell_1$-norm is the soft-thresholding operator,
\vspace{-2pt}
\begin{equation}
    \left(\mathcal{T}_{(1/\rho)\|{\cdot}\|_1}({\bf z})\right)[m,n]
    =\left(1-\frac{1/\rho}{|z[m,n]|}\right)_+z[m,n],
    \label{eq:soft}
    \vspace{-2pt}
\end{equation}
where ${\bf z}\in \mathbb{C}^{M{\times}N}$, and $(\cdot)_+={\text{max}}(\:{\cdot}\;,\,0)$.
In order to adjust bin-wise regularization strength, the weighted $\ell_1$-norm can be introduced.
For example, the parabola weight $w[m,n] = (m+1)^2/M^2$ was introduced in \cite{PWL1}, and its effectiveness was confirmed.
Besides this, some generalized thresholding operators \cite{SS,HVA,Social,shrinkage} and heuristic thresholding \cite{ASPADE} have been also utilized for audio declipping.
 
%Such a thresholding operator is usually a bin-wise process in which inputs whose absolute value is smaller than a certain value are replaced by 0, and others are reduced.
%Interestingly, several generalized thresholding functions \cite{HVA} have been proposed that are defined directly in the form of thresholding functions without corresponding sparsity-promoting functions.
%Although they deviate from the framework: the proximity operator of a sparsity-promoting function, they have been used in various applications including audio declipping due to their flexibility to promote sparsity \cite{Social,shrinkage}.
%Some methods \cite{ASPADE,PWL1} also investigated on the way to promote sparsity, and they revealed its importance in audio declipping.

\vspace{-5pt}
\subsection{Inappropriateness of soft-thresholding}
\label{ssec:soft}
\vspace{-1pt}

\begin{figure}
	\centering
	\vspace{-4pt}
	\includegraphics[width=0.99\columnwidth]{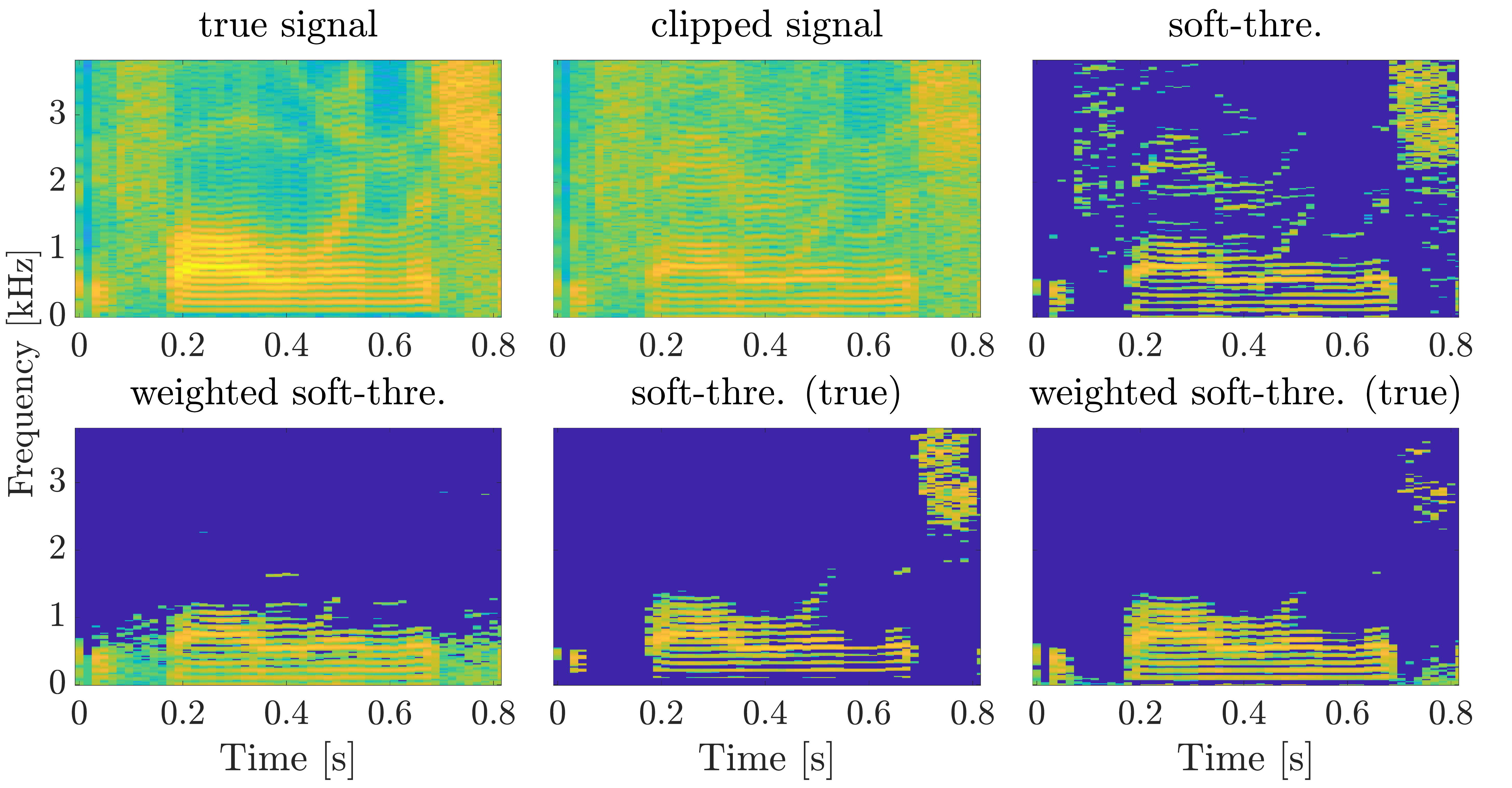}
	\vspace{-0.4cm}
	\caption{Amplitude spectrograms with various processing. $\tau$ was fixed to 0.01, and $\rho$ was adjusted for ease of viewing. The spaectrograms with (true) were processed by the soft-thresholding operator whose denominator was the power spectrogram of the true signal.}
	\label{fig:mask}
\end{figure}

The conventional thresholding operators have some disadvantages due to their uniform and data-irrelevant nature.
Let us explain it using Fig.~\ref{fig:mask}.
The upper left and  center figures are the spectrograms of the true and a clipped signal, respectively.
Clipping generated some extra harmonic components.
The upper right figure is the spectrogram processed by Eq.~\eqref{eq:soft}.
\begin{comment}
It cannot remove the extra components properly.
\end{comment}
The extra components were not removed properly.
The spectrogram processed by the parabola-weighted soft-thresholding operator \cite{PWL1} is given in the lower left.
Although the weights protected the low-frequency part, the high-frequency components were severely removed.
These disadvantages are due to the fact that these operators are unrelated to the properties of the data.

If the true signal is available, we can promote sparsity more reasonably.
The lower center figure is the spectrogram of the clipped signal processed by the soft-thresholding operator in Eq.~\eqref{eq:soft} whose denominator $|z[m,n]|$ was replaced by the power spectrogram of the true signal $|(\mathcal{G}\tilde{\bf y})[m,n]|^2$.
The figure shows that extra harmonic components were properly removed.
The lower right figure was processed by the true-signal-based soft thresholding operator with the parabola weight.
The necessary parts remained even in high frequency.
Therefore, if the true signals are available, we can properly remove the extra harmonic components by the thresholding operator.%

\vspace{-5pt}
\subsection{APPLADE: Adjustable Plug-and-PLay Audio DEclipper}
\label{ssec:applade}
\vspace{-1pt}
    
Here, we propose APPLADE to utilize the properties of data learned by a DNN.
As we show in the previous subsection, the conventional thresholding uniformly promote sparsity, and it can lead to inappropriate results.
As one solution to this, we introduced thresholding using the true signal, but of course it is not available.
This unrealistic thresholding can be approximated using a DNN ${\mathcal{F}_\theta}$ that estimates the magnitude spectrogram of the true signal from that of a clipped signal.
We use ${\mathcal{F}_\theta}$ in the weighted soft-thresholding operator as
\vspace{-2pt}
\begin{equation}
    ({\mathcal T}_\theta({\bf z}))[m,n] = \left(1 - \frac{\lambda\,w[m,n]}{(({\mathcal{F}_\theta(|{\bf z}|)})[m,n] + \epsilon)^2}\right)_+z[m,n],
    \label{eq:propsoft}
    \vspace{-2pt}
\end{equation}
where $\bf w$ is the parabola weight \cite{PWL1}, $\lambda$ is a parameter to adjust the strength of thresholding, and $\epsilon$ is a small constant for numerical stability.
We use this operator in place of $\text{prox}_{(1/\rho){\mathcal{S}}}$ of the ADMM in Eq.~\eqref{eq:2}, and call this new algorithm APPLADE.
The entire algorithm is shown in Alg.~\ref{alg:applade}\footnote{Code publicly available at \url{https://doi.org/hgzf}.}, where $\mathcal{F}_\theta$ is applied repeatedly in the 5th line.
In general, careless use of a DNN in an iterative algorithm can be very unstable (as will be demonstrated by experiments).
Thus, we propose the operator in Eq.~\eqref{eq:propsoft} to reduce such undesirable effect.

\begin{algorithm}[t]
\caption{APPLADE}
\label{alg:applade}
\begin{algorithmic}[1]
\STATE \textbf{Input:} ${\bf x}^{[0]}$, ${\bf v}^{[0]}$, ${\bf u}^{[0]}$
\STATE \textbf{Output:} ${\bf x}^{[K]}$
\FOR{$k=0,1,\dots,K-1$}
\STATE ${\bf x}^{[k+1]} = \Pi_\Gamma(\mathcal{G}^*({\bf v}^{[k]} - {\bf u}^{[k]}))$
\STATE ${\bf v}^{[k+1]} = \mathcal{T}_\theta(\mathcal{G}{\bf x}^{[k+1]} + {\bf u}^{[k]})$
\STATE ${\bf u}^{[k+1]} = {\bf u}^{[k]} + \mathcal{G}{\bf x}^{[k+1]} - {\bf v}^{[k+1]}$
\ENDFOR
\end{algorithmic}
\end{algorithm}

The proposed method has five advantages.
First, $\mathcal{T}_\theta$ can promote sparsity based on the features of data.
This leads to more appropriate thresholding for audio declipping than the conventional ones.
Second, any DNN can be used as $\mathcal{F}_\theta$. 
Third, training of the DNN is independent of the iterative algorithm.
Fourth, the proposed method inherits
adjustability from the sparsity-based methods.
Finally, the clipping consistency is considered during processing, i.e., the proposed method does not alter the unclipped elements.

\vspace{-2pt}
\section{Experiments and Results}
\label{sec:results}
\vspace{-2pt}

\vspace{-3pt}
\subsection{Training}
\label{ssec:training}
\vspace{-1pt}

LIBRI speech corpus \cite{LIBRI} was utilized for training.
We used 5323 clean speech signals (5300 for training, 23 for validation), and 16384 samples of a voiced part were extracted from each signal (about 1 s at sampling frequency of 16 kHz).
All data were peak-normalized.
They were corrupted by clipping, where $\tau$ was set according to signal-to-distortion ratio (SDR), $\text{SDR}({\bf{x}},{\bf{y}})=20\log_{10}{\|{\bf{x}}\|_2}/{\|{\bf x} - {\bf y}\|_2}$.
The input SDR was randomly drawn from the uniform distribution in the interval $[1, 10]$.

\begin{figure}[t]
	\centering
	\includegraphics[width=0.99\columnwidth]{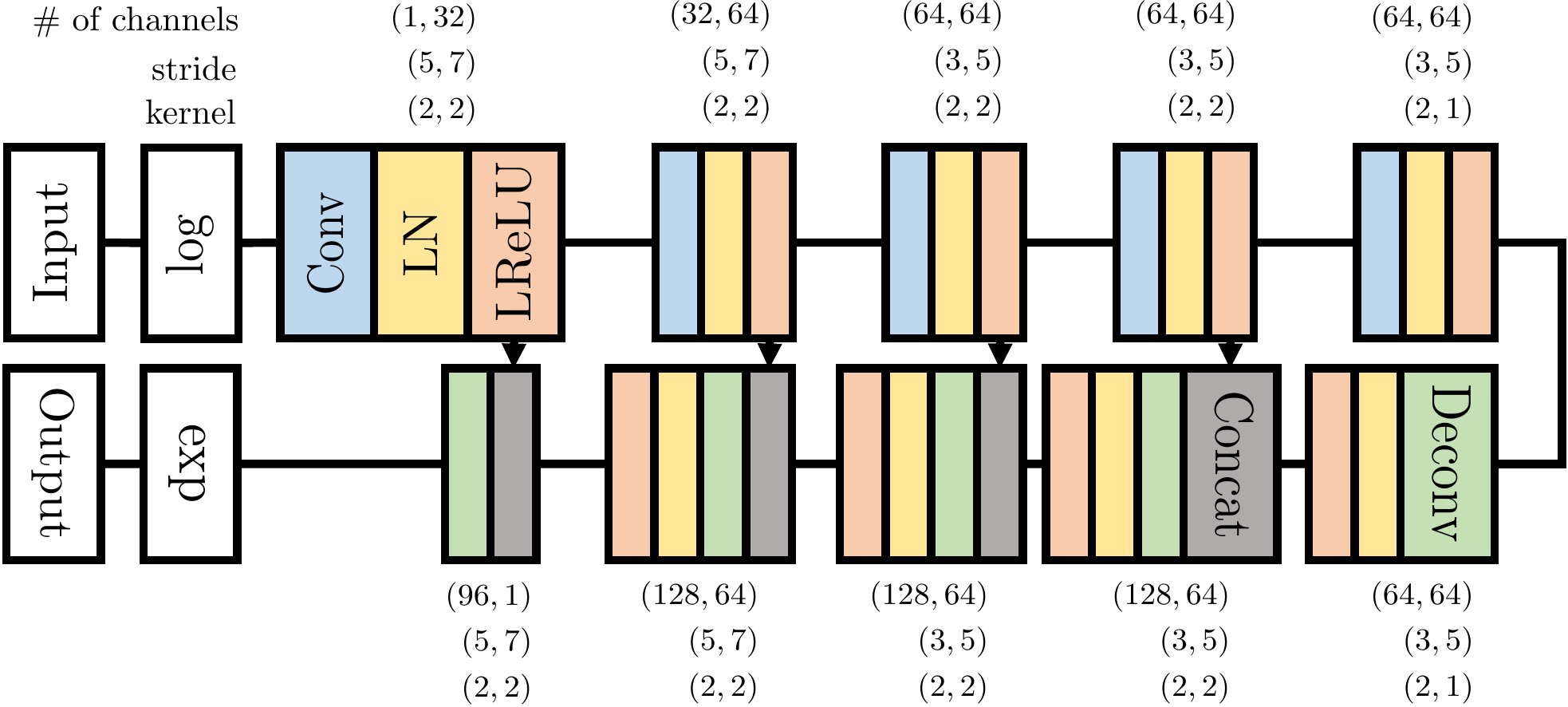}
	\vspace{-0.4cm}
	\caption{Network used in our experiments. ''Conv", ''Deconv" ''LN", ``LReLU'', and ''Concat" stand for convolution, deconvolution, layer normalization, leaky ReLU, and concatenation, respectively.}
	\label{fig:DNN}
\end{figure}

The structure of the DNN used as $\mathcal{F}_\theta$ is shown in Fig.~\ref{fig:DNN}.
The U-Net architecture was designed based on \cite{unet,unet2}.
Since we use $\mathcal{F}_\theta$ iteratively as in Alg.~\ref{alg:applade}, using a too large network will lead to an increase in computation time.
Thus, we modified its structure to avoid that.
The number of parameters is about one-eleventh of that in TF-UNet \cite{TTF} (856,033 vs. 9,711,361).
The DNN 
\begin{comment}
in Fig.~\ref{fig:DNN} 
\end{comment}
was trained 200 epochs with the adam optimizer \cite{Adam} with a batch size of 4, a learning rate of 0.0001, and decay rates of $\beta_1=0.9$ and $\beta_2 = 0.999$.
75 \% overlapped 1024-point-long-Hann window was used for DGT.
The input and output features were magnitude spectrograms sized $512\times64$ by removing the highest (Nyquist) frequency bin.
When the output was used in Eq.~\eqref{eq:propsoft}, we padded zeros so that it has the original size $513\times64$.
The scale of the leaky ReLU was 0.01.
The loss function was the time-domain mean-squared-error (MSE) computed via inverse DGT.
The loss was calculated only in clipped parts $H$ and $L$.
After the training of DNN, $\lambda$ was determined using the validation data. By curve fitting, $\lambda=30\,p$ was obtained, where $p = (|H|+|L|)/T$ is the ratio of the number of clipped samples.

\vspace{-3pt}
\subsection{Testing}
\label{ssec:testing}
\vspace{-0.5pt}

For testing, 200 speech signals \cite{timit} sampled at 16 kHz from TIMIT corpus were used.
Clipping threshold $\tau$ was set according to SDR (1, 3, 5, 10, 15 dB). 
All data were cut out into 16384 samples and peak-normalized.
The initial values were set as follows: ${\bf x}^{[0]}$ was set to an observed clipped signal, ${\bf v}^{[0]} = \mathcal{G}{\bf x}^{[0]}$, and ${\bf u}^{[0]} = {\bf 0}$.
$\epsilon$ was set to $10^{-6}$, and the number of iteration $K$ was set to 200.
We used $\Delta\text{SDR} = \text{SDR}(\hat{\bf y},{\tilde{\bf x}}) - \text{SDR}(\hat{\bf y},{\bf y})$ and $\Delta$PESQ, which is improvement of PESQ \cite{PESQ}, for evaluation of quality of restored signals.

\begin{figure}[t]
	\centering
	\includegraphics[width=0.99\columnwidth]{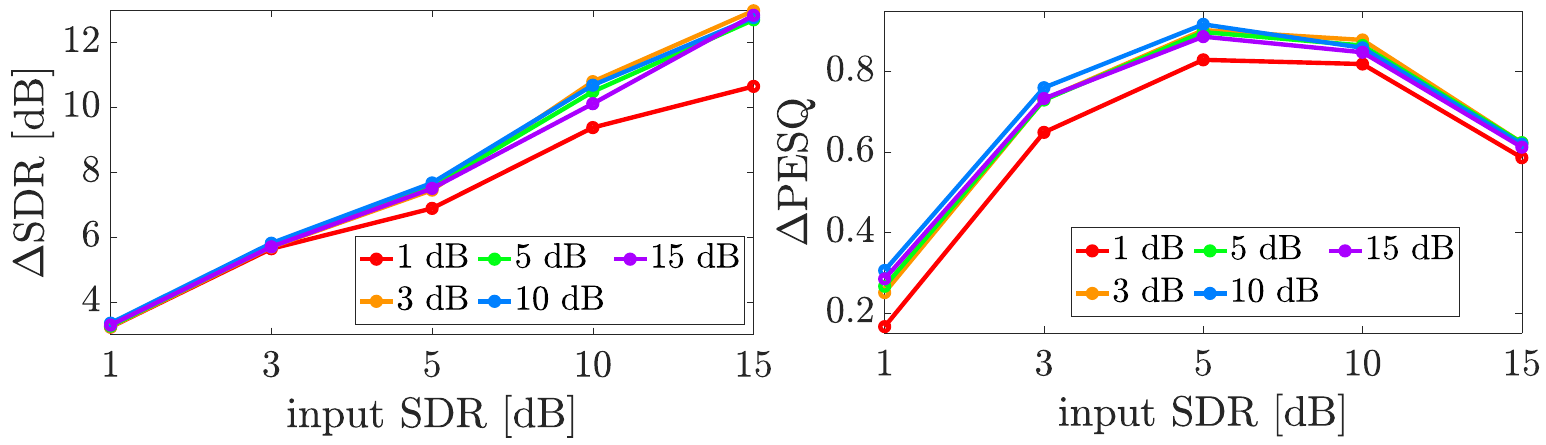}
	\vspace{-0.3cm}
	\caption{Median of $\Delta$SDR and $\Delta$PESQ of APPLADE when there is the mismatch of clipping level (input SDR) between the training and test data. The legend indicates the input SDR of training data.}
	\label{fig:mismatch}
	\vspace{-0.1cm}
\end{figure}

\vspace{-3pt}
\subsection{Experiment 1: Mismatch between training and test data}
\label{ssec:exp1}
\vspace{-1pt}

We conducted an experiment to see how the proposed method behaves for unseen data.
During training, all training data were clipped by the same clipping level (input SDR was either 1, 3, 5, 10 or 15 dB).
Then, the proposed method was tested for all clipping levels (i.e., 4 out of 5 conditions were unseen).
The results are shown in Fig.~\ref{fig:mismatch}.
The mismatch between input SDR during training and testing did not result in a significant performance degradation.
Therefore, even for unseen clipping levels, the proposed method will work properly with DNNs trained with an appropriate range of input SDR.

\begin{figure}[t]
	\centering
	\includegraphics[width=0.99\columnwidth]{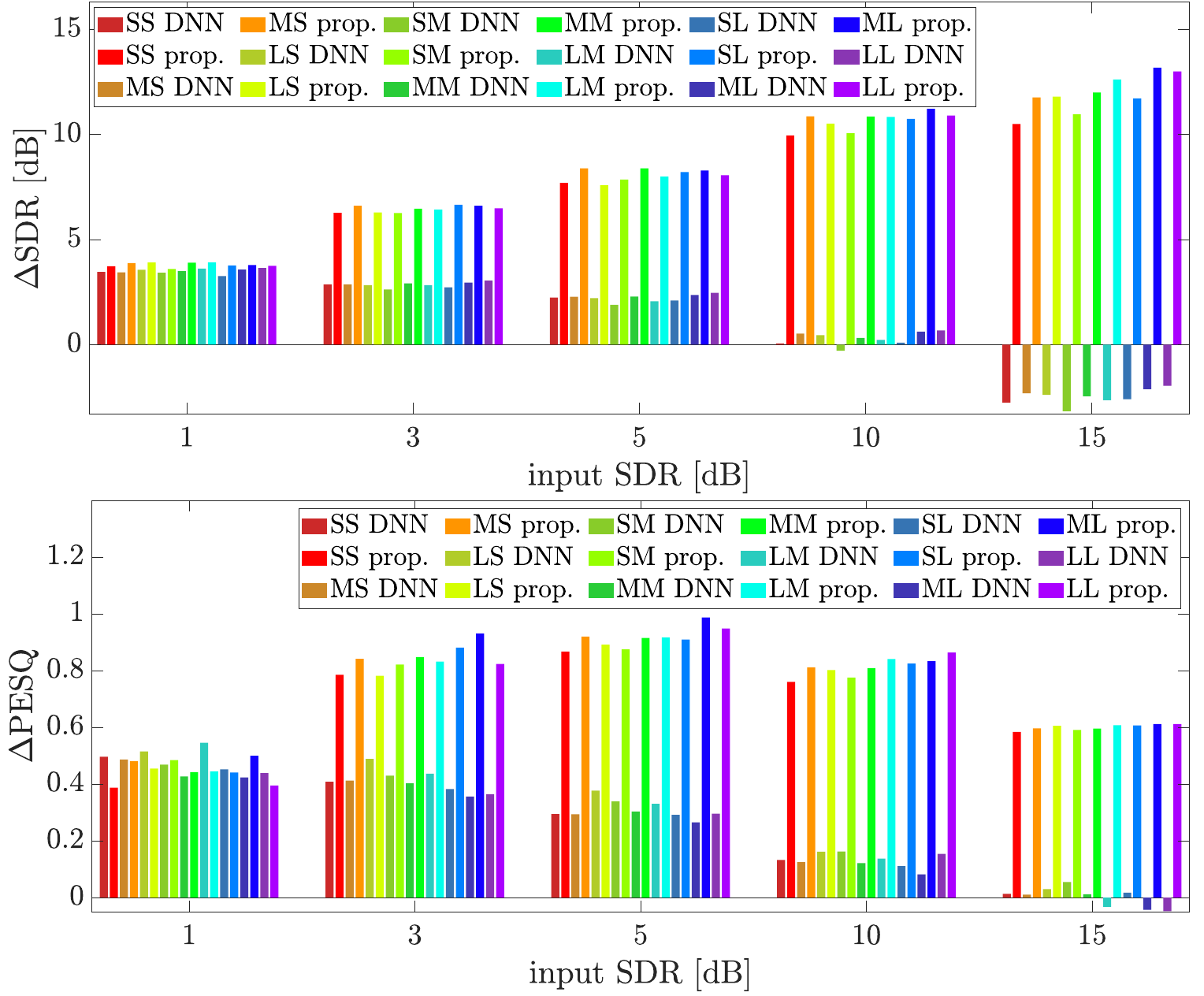}
	\vspace{-0.3cm}
	\caption{Median of $\Delta$SDR and $\Delta$PESQ when the number of DNN parameters was varied. The first letter indicates the number of blocks (S: 3, M: 4, L: 5). The second letter indicates the number of channels of Conv/Deconv (S: 32, M:64, L: 128). ``prop.'' means that the same DNN was inserted and used in the proposed algorithm.}
	\label{fig:structure}
	\vspace{-0.2cm}
\end{figure}

\vspace{-3pt}
\subsection{Experiment 2: Effect of the number of DNN parameters}
\label{ssec:exp2}
\vspace{-1pt}

Next, we conducted an experiment to see the effect of the number of DNN parameters on the performance of APPLADE.
To naturally change the number of parameters, we used a special DNN only for this experiment.
The DNN had the U-Net architecture, which has the same number of blocks for the encoder and decoder.
The number of blocks were chosen from 3, 4, 5.
Each block was composed of Conv/Deconv (kernel size: $(5,7)$, stride size: $(2,2)$), LN, and LReLU, but the last block of the decoder was Deconv only.
The number of channels of Conv/Deconv were chosen from 32, 64, 128.
Hence, 9 DNNs (determined by the combination of the numbers of blocks and channels) were trained as in Sec.~\ref{ssec:training}.
To see the effect of the iterative algorithm, the trained DNNs were used \textit{with} and \textit{without} the proposed algorithm.

Fig.~\ref{fig:structure} shows the results.
When using the DNNs without the proposed algorithm (darker colors), the performance of all 9 DNNs was poor.
In contrast, APPLADE using them as $\mathcal{F}_\theta$ in Eq.~\eqref{eq:propsoft} (brighter colors) performed notably better.
Interestingly, the performance of APPLADE was not greatly affected by the number of parameters.
Also, the performance of APPLADE seems unrelated to the performance of the DNN itself.
These results indicates that the proposed method can absorb the difference of the DNNs to some extent, which is a desirable property because choice of a DNN is not restricted.

\vspace{-3pt}
\subsection{Experiment 3: Comparison with other methods}
\label{ssec:exp3}
\vspace{-1pt}

The proposed method was compared with some other methods.
As sparsity-based methods, consistent-IHT (IHT) \cite{IHT}, ASPADE \cite{ASPADE}, Social-Sparsity with Persistent Empirical Wiener (SS PEW) \cite{SS}, and Parabola-Weighted $\ell_1$ minimization (PW$\ell_1$) \cite{PWL1} were performed.
The parameters and other detailed settings were taken from the original paper or the summary paper \cite{summary} to suit for the speech signals sampled at 16 kHz.
As a DNN-based method, T-UNet \cite{TTF} $\mathcal{F}^{\text{T-UNet}}_\theta$ approximating $\mathcal{F}^{\text{TD}}({\bf y}) = \tilde{\bf y}$ was applied.
The number of parameters of T-UNet was 11,283,585 \cite{TTF}, and this was about 13 times larger than that of the DNN used in the proposed method (Fig.~\ref{fig:DNN}).
In addition, the naive PnP method that uses a Gaussain denoising DNN\footnote{We changed the output of Fig.~\ref{fig:DNN} to the sigmoid function and trained it to remove 1 to 10 dB of time-domain white Gaussian noise by T-F masking.}
as $\text{prox}_{(1/\rho){\mathcal{S}}}$ (naive PnP) was also performed.

\begin{figure}[t]
    \vspace{-2pt}
	\centering
	\includegraphics[width=0.99\columnwidth]{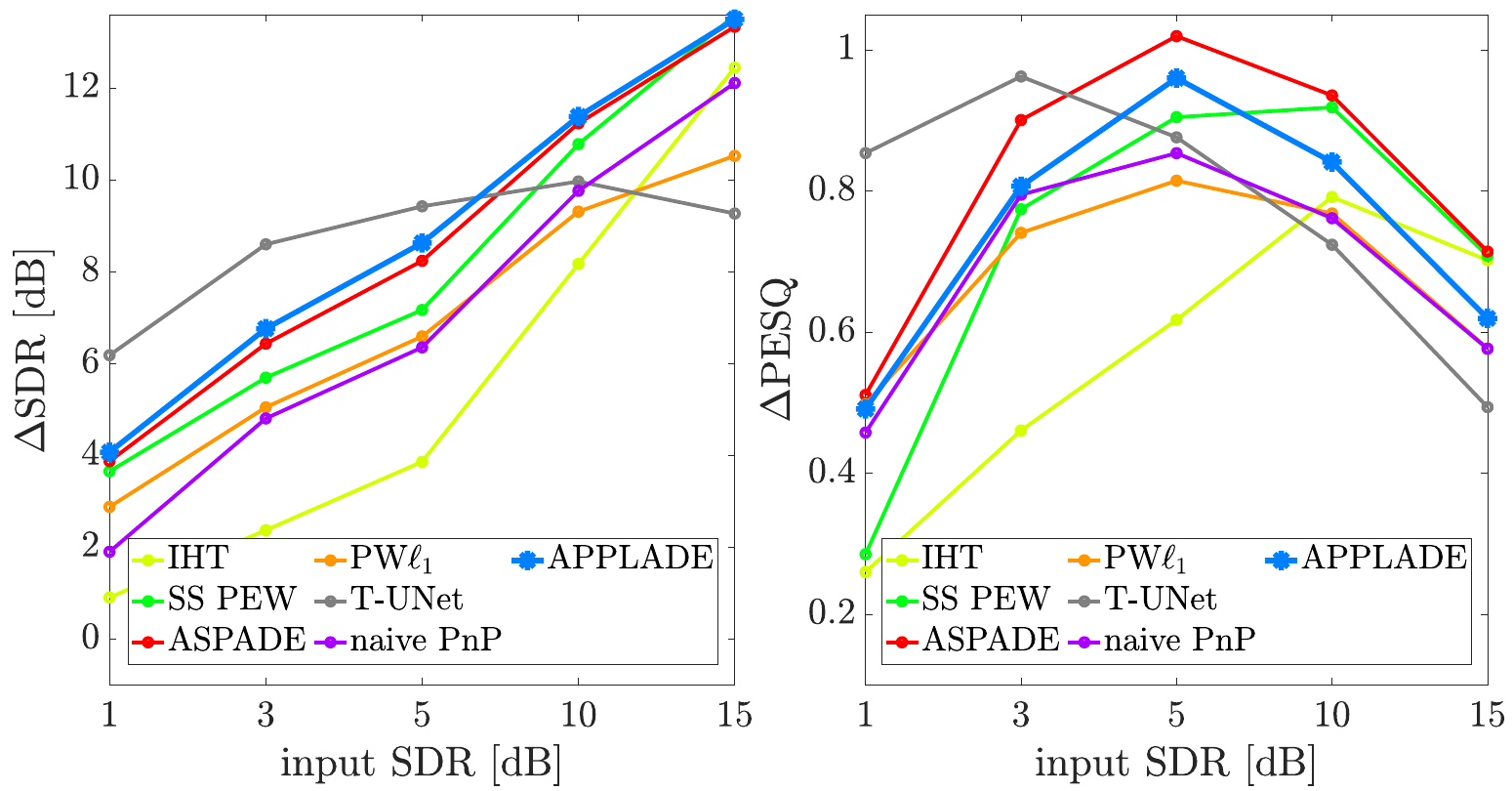}
	\vspace{-0.3cm}
	\caption{Median of $\Delta$SDR and $\Delta$PESQ for each input SDR.}
	\label{fig:comp}
	\vspace{-0.1cm}
\end{figure}

Fig.~\ref{fig:comp} shows the results.
First of all, let us focus on  APPLADE (blue) and PW$\ell_1$ (orange).
The  difference between them is whether the thresholding operator uses the DNN or not.
By effectively using the DNN, $\Delta$SDR was improved by about 2 dB on average.
Compared to other sparsity-based methods, the proposed method restored clipped signals best in terms of $\Delta$SDR and as well as ASPADE (red) and SS PEW (green) in terms of $\Delta$PESQ.
T-UNet (gray) performed better in low input SDR (i.e., when the clipping was severe), but not in high input SDR.
This should be because it cannot consider the clipping consistency, and hence T-UNet does not have information about the position of samples that must be restored.
On the contrary, the sparsiy-based methods and APPLADE can take advantage of the clipping consistency by the projection in Eqs.~\eqref{eq:1} and \eqref{eq:pj}.

\begin{figure}[t]
	\centering
	\includegraphics[width=0.99\columnwidth]{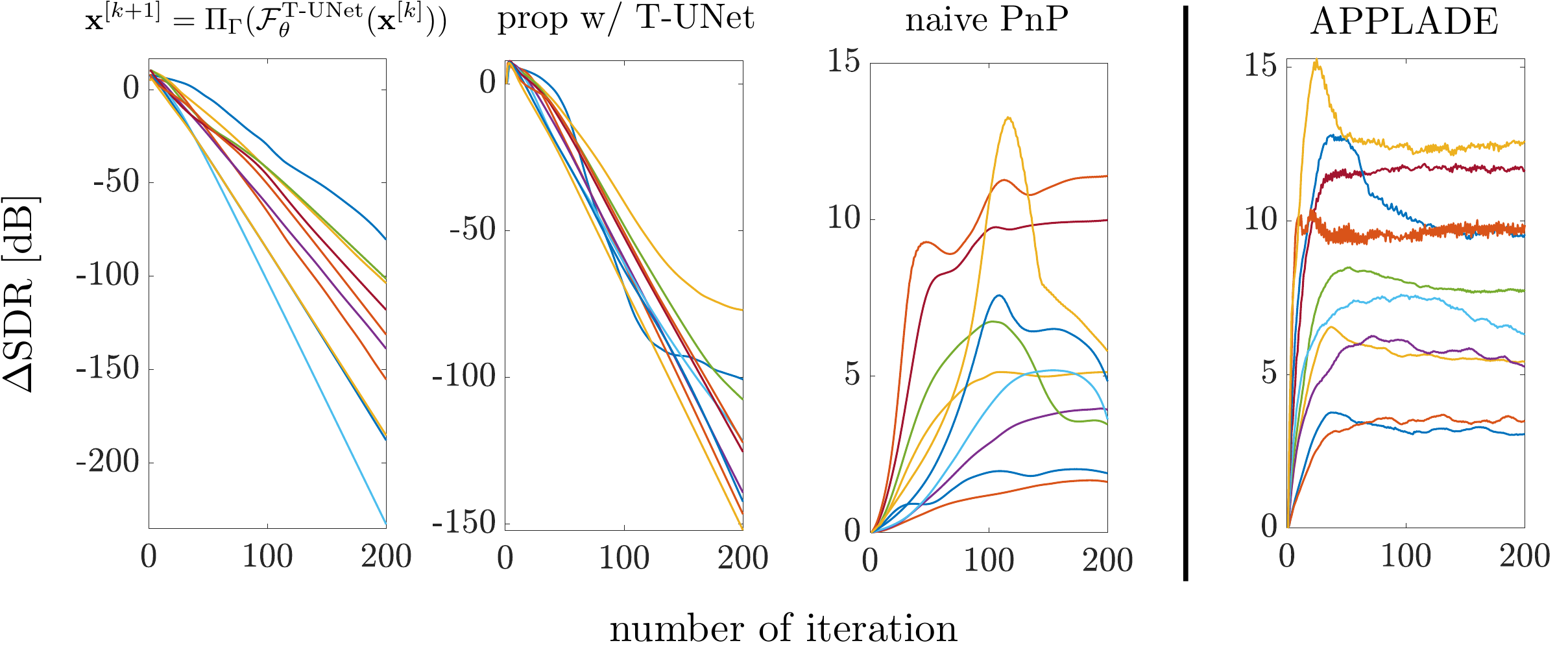}
	\vspace{-0.3cm}
	\caption{$\Delta$SDR for each iteration. Results for 10 signals are shown.}
	\label{fig:periter}
	\vspace{-0.1cm}
\end{figure}

To demonstrate the appropriateness of the proposed method, it was compared with some other combinations of the DNN and an iterative algorithm.
Fig.~\ref{fig:periter} shows $\Delta$SDR of 10 signals for each iteration.
Two methods were implemented in addition to naive PnP: a simple iteration of T-UNet and projection, and the ADMM algorithm using T-UNet without the proposed operator in Eq.~\eqref{eq:propsoft}.
Although T-UNet itself performs well as in Fig.~\ref{fig:comp}, the two simple methods using T-UNet on the left completely failed in audio declipping.
The naive PnP obtained smooth curves, but its performance was unstable.
Only APPLADE was able to achieve reasonable results for all signals.
These results indicate that the proposed operator in Eq.~\eqref{eq:propsoft} is essential for successfully perform audio declipping.

Table~\ref{tab:time} shows execution time of each method computed by Intel Core i9-10900K (or RTX 2060 for APPLADE w/GPU).
Since the proposed method can achieve a good performance with less iterations, the total time ($t_2$) was notably smaller than the other methods.

\begin{table}[t]
\vspace{-6pt}
\centering
\caption{Execution time per iteration $t_1$ and execution time $t_2$ necessary for reaching 95 \%  of $\Delta$SDR in Fig.~\ref{fig:comp}.}
\vspace{2pt}
\begin{threeparttable}[h]
\scalebox{0.92}{
\begin{tabular}{c|C{30pt}C{33pt}C{40pt}C{28pt}C{45pt}C{45pt}}
    \toprule
     & IHT$^\ast\!\!$ & SS PEW & ASPADE$^\ast\!\!$ & PW$\ell_1\!$ & \begin{tabular}{@{}c@{}}
     APPLADE \\ w/ CPU 
     \end{tabular} &
     \begin{tabular}{@{}c@{}}
     APPLADE \\ w/ GPU
     \end{tabular}\\
    \midrule
    $t_1$ [s] & $0.031$ & $0.037$ & $0.030$ & $0.037$ & $0.025$ & $0.023$ \\
    $t_2$ [s] & $5.2$ & $79$ & $8.5$ & $0.91$ & $0.72$ & $0.67$\\
    \bottomrule
\end{tabular}
}
\begin{tablenotes}
\item[$\ast$]\footnotesize Since ASPADE and IHT were performed window-wise, we multiplied \\  $t_1$ and $t_2$ for one window by the number of windows.
\end{tablenotes}
\end{threeparttable}
\label{tab:time}
\vspace{-12pt}
\end{table}

\vspace{-6pt}
\section{Conclusions}
\label{sec:conc}
\vspace{-2pt}

In this paper, we proposed the PnP audio declipping method named APPLADE. 
A DNN is embedded in the ADMM algorithm and helps in better thresholding for audio declipping.
Our experiments showed that APPLADE was robust to variations in the number of DNN parameters and unseen data.
Moreover, APPLADE was found to reach higher performance in less time than the other sparsity-based methods and to be more stable than the other possible methods combining a DNN and iteration.
Future work will be on theoretical guarantees of convergence and applying to other clipping models.

\vfill%\pagebreak
\clearpage

% References should be produced using the bibtex program from suitable
% BiBTeX files (here: strings, refs, manuals). The IEEEbib.bst bibliography
% style file from IEEE produces unsorted bibliography list.
% -------------------------------------------------------------------------
\bibliographystyle{IEEEbib}
\bibliography{refs}

\begin{thebibliography}{10}

\bibitem{q2}
C.-T. Tan and B.C.J. Moore,
\newblock ``Perception of nonlinear distortion by hearing-impaired people,''
\newblock {\em Int. J. Audiol.}, vol. 47, no. 5, pp. 246--256, 2008.

\bibitem{q1}
K.H. Arehart, J.M. Kates, and M.C. Anderson,
\newblock ``Effects of noise, nonlinear processing, and linear filtering on
  perceived music quality,''
\newblock {\em Int. J. Audiol.}, vol. 50, no. 3, pp. 177--190, 2011.

\bibitem{r2}
J.~{Malek},
\newblock ``Blind compensation of memoryless nonlinear distortions in sparse
  signals,''
\newblock in {\em 21st Eur. Signal Process. Conf. (EUSIPCO)}, 2013, pp. 1--5.

\bibitem{r1}
Y.~Tachioka, T.~Narita, and J.~Ishii,
\newblock ``Speech recognition performance estimation for clipped speech based
  on objective measures,''
\newblock {\em Acoust. Sci. Technol.}, vol. 35, no. 6, pp. 324--326, 2014.

\bibitem{IHT}
S.~Kitic, L.~Jacques, N.~Madhu, M.P. Hopwood, A.~Spriet, and
  C.~De~Vleeschouwer,
\newblock ``Consistent iterative hard thresholding for signal declipping,''
\newblock in {\em IEEE Int. Conf. Acoust. Speech Signal Process. (ICASSP)},
  2013, pp. 5939--5943.

\bibitem{SS}
K.~Siedenburg, M.~Kowalski, and M.~Dörfler,
\newblock ``Audio declipping with social sparsity,''
\newblock in {\em IEEE Int. Conf. Acoust. Speech Signal Process. (ICASSP)},
  2014, pp. 1577--1581.

\bibitem{ASPADE}
P~Z\'{a}vi\v{s}ka, P~Rajmic, O.~Mokr\'{y}, and Z.~Pr\r{u}\v{s}a,
\newblock ``A proper version of synthesis-based sparse audio declipper,''
\newblock in {\em IEEE Int. Conf. Acoust. Speech Signal Process. (ICASSP)},
  2019, pp. 591--595.

\bibitem{PWL1}
P.~Z\'{a}vi\v{s}ka, P.~Rajmic, and J.~Schimmel,
\newblock ``Psychoacoustically motivated audio declipping based on weighted
  $\ell_1$ minimization,''
\newblock in {\em Int. Conf. Telecommun. Signal Process. (TSP)}, 2019, pp.
  338--342.

\bibitem{multi}
S.~Emura and N.~Harada,
\newblock ``An extension of sparse audio declipper to multiple measurement
  vectors,''
\newblock in {\em IEEE Int. Conf. Acoust. Speech Signal Process. (ICASSP)},
  2021, pp. 686--690.

\bibitem{summary}
C.~Gaultier, S.~Kiti\'{c}, R.~Gribonval, and N.~Bertin,
\newblock ``Sparsity-based audio declipping methods: Selected overview, new
  algorithms, and large-scale evaluation,''
\newblock {\em IEEE/ACM Trans. Audio Speech Lang. Process.}, vol. 29, pp.
  1174--1187, 2021.

\bibitem{DNN3}
F.~Bie, D.~Wang, J.~Wang, and T.F. Zheng,
\newblock ``Detection and reconstruction of clipped speech in speaker
  recognition,''
\newblock {\em Speech Commun.}, vol. 72, 07 2015.

\bibitem{DF}
W.~Mack and E.A.P. Habets,
\newblock ``Declipping speech using deep filtering,''
\newblock in {\em IEEE Workshop Appl. Signal Process. Audio Acoust. (WASPAA)},
  2019, pp. 200--204.

\bibitem{TTF}
A.A. Nair and K.~Koishida,
\newblock ``Cascaded time + time-frequency {U}net for speech enhancement:
  Jointly addressing clipping, codec distortions, and gaps,''
\newblock in {\em IEEE Int. Conf. Acoust. Speech Signal Process. (ICASSP)},
  2021, pp. 7153--7157.

\bibitem{blstm}
S.~Hochreiter and J.~Schmidhuber,
\newblock ``Long short-term memory,''
\newblock {\em Neural Comput.}, vol. 9, no. 8, pp. 1735–1780, Nov. 1997.

\bibitem{unet}
O.~Ronneberger, P.~Fischer, and T.~Brox,
\newblock ``U-{N}et: Convolutional networks for biomedical image
  segmentation,''
\newblock in {\em Med. Image Comput. Comput.-Assist. Interv. (MICCAI)}. 2015,
  pp. 234--241, Springer International Publishing.

\bibitem{PnP1}
S.V. Venkatakrishnan, C.A. Bouman, and B.~Wohlberg,
\newblock ``Plug-and-play priors for model based reconstruction,''
\newblock in {\em IEEE Glob. Conf. Signal Inf. Process.}, 2013, pp. 945--948.

\bibitem{PnP2}
U.S. Kamilov, H.~Mansour, and B.~Wohlberg,
\newblock ``A plug-and-play priors approach for solving nonlinear imaging
  inverse problems,''
\newblock {\em IEEE Signal Process. Lett.}, vol. 24, no. 12, pp. 1872--1876,
  2017.

\bibitem{PnP3}
S.H. Chan, X.~Wang, and O.A. Elgendy,
\newblock ``Plug-and-play {ADMM} for image restoration: Fixed-point convergence
  and applications,''
\newblock {\em IEEE Trans. Comput. Imaging}, vol. 3, no. 1, pp. 84--98, 2017.

\bibitem{PnP4}
T.~Meinhardt, M.~Moeller, C.~Hazirbas, and D.~Cremers,
\newblock ``Learning proximal operators: Using denoising networks for
  regularizing inverse imaging problems,''
\newblock in {\em IEEE Int. Conf. Comput. Vis. (ICCV)}, 2017, pp. 1799--1808.

\bibitem{PnP5}
M~Terris, A~Repetti, J.-C. Pesquet, and Y.~Wiaux,
\newblock ``Building firmly nonexpansive convolutional neural networks,''
\newblock in {\em IEEE Int. Conf. Acoust. Speech Signal Process. (ICASSP)},
  2020, pp. 8658--8662.

\bibitem{PnP6}
J.-C. Pesquet, A.~Repetti, M.~Terris, and Y.~Wiaux,
\newblock ``Learning maximally monotone operators for image recovery,''
\newblock {\em SIAM J. Imaging Sci.}, vol. 14, no. 3, pp. 1206--1237, 2021.

\bibitem{proximal}
N.~Parikh and S.~Boyd,
\newblock {\em Proximal Algorithms},
\newblock Now Publishers Inc., 2014.

\bibitem{playwdual}
N.~{Komodakis} and J.-C. {Pesquet},
\newblock ``Playing with duality: An overview of recent primal-dual approaches
  for solving large-scale optimization problems,''
\newblock {\em IEEE Signal Process. Mag.}, vol. 32, no. 6, pp. 31--54, 2015.

\bibitem{monotone}
E.~Ryu and S.~Boyd,
\newblock ``A primer on monotone operator methods survey,''
\newblock {\em Appl. comput. math.}, vol. 15, pp. 3--43, 01 2016.

\bibitem{HVA}
K.~Yatabe and D.~Kitamura,
\newblock ``Determined {BSS} based on time-frequency masking and its
  application to harmonic vector analysis,''
\newblock {\em IEEE/ACM Trans. Audio Speech Lang. Process.}, vol. 29, pp.
  1609--1625, 2021.

\bibitem{Social}
M.~Kowalski, K.~Siedenburg, and M.~D\"{o}rfler,
\newblock ``Social sparsity! neighborhood systems enrich structured shrinkage
  operators,''
\newblock {\em IEEE Trans. Signal Process.}, vol. 61, no. 10, pp. 2498--2511,
  2013.

\bibitem{shrinkage}
R.~Chartrand,
\newblock ``Shrinkage mappings and their induced penalty functions,''
\newblock in {\em IEEE Int. Conf. Acoust. Speech Signal Process. (ICASSP)},
  2014, pp. 1026--1029.

\bibitem{LIBRI}
V.~Panayotov, G.~Chen, D.~Povey, and S.~Khudanpur,
\newblock ``Librispeech: An {ASR} corpus based on public domain audio books,''
\newblock in {\em IEEE Int. Conf. Acoust. Speech Signal Process. (ICASSP)},
  2015, pp. 5206--5210.

\bibitem{unet2}
D.~Takeuchi, K.~Yatabe, Y.~Koizumi, Y.~Oikawa, and N.~Harada,
\newblock ``Effect of spectrogram resolution on deep-neural-network-based
  speech enhancement,''
\newblock {\em Acoust. Sci. Technol.}, vol. 41, no. 5, pp. 769--775, 2020.

\bibitem{Adam}
D.P. {Kingma} and J.L. {Ba},
\newblock ``Adam: A method for stochastic optimization,''
\newblock in {\em Proc. IEEE Int. Conf. on Learn. Represent. (ICLR)}, 2015.

\bibitem{timit}
P.~Mowlaee, J.~Kulmer, J.~Stahl, and F.~Mayer,
\newblock {\em Single Channel Phase-Aware Signal Processing in Speech
  Communication: Theory and Practice},
\newblock Hoboken, NJ, USA: Wiley, 11 2016.

\bibitem{PESQ}
A.W. Rix, J.G. Beerends, M.P. Hollier, and A.P. Hekstra,
\newblock ``Perceptual evaluation of speech quality ({PESQ})-a new method for
  speech quality assessment of telephone networks and codecs,''
\newblock in {\em IEEE Int. Conf. Acoust. Speech Signal Process. (ICASSP)},
  2001, vol.~2, pp. 749--752.

\end{thebibliography}

\end{document}